\RequirePackage{fix-cm}
\documentclass[smallextended]{svjour3}       % onecolumn (second format)
\smartqed  % flush right qed marks, e.g. at end of proof
\usepackage{graphicx}
\usepackage{xcolor}
\usepackage{booktabs}
\usepackage{verbatim}
\usepackage{color, colortbl}
\usepackage{url}

\begin{document}

\title{FootApp: an AI-Powered System for Football Match Annotation%\thanks{Grants or other notes
}


\author{Silvio Barra \and Salvatore M. Carta \and Alessandro Giuliani \and Alessia Pisu \and Alessandro Sebastian Podda \and Daniele Riboni}


\institute{S. Barra \at
              Department of Information Technology and Electric Engineering\\ 
              University of Naples ``Federico II", Via Claudio 21, 80125 Napoli (Italy)\\
              %\email{fauthor@example.com}           %  \\
           \and
           S. M. Carta \and A. Giuliani \and A. Pisu \and A.S. Podda \and D. Riboni \at
              Department of Mathematics and Computer Science\\ 
              University of Cagliari, Via Ospedale 72, 09124 Cagliari (Italy)\\
              %\email{riboni@unica.it}
}

\date{Received: date / Accepted: date}

\maketitle

\begin{abstract}
In the last years, scientific and industrial research has experienced a growing interest in acquiring large annotated data sets to train artificial intelligence algorithms for tackling problems in different domains. In this context, we have observed that even the market for football data has substantially grown. The analysis of football matches relies on the annotation of both individual players' and team actions, as well as the athletic performance of players. Consequently, annotating football events at a fine-grained level is a very expensive and error-prone task. Most existing semi-automatic tools for football match annotation rely on cameras and computer vision. However, those tools fall short in capturing team dynamics, and in extracting data of players who are not visible in the camera frame. To address these issues, in this manuscript we present FootApp, an AI-based system for football match annotation. First, our system relies on an advanced and mixed user interface that exploits both vocal and touch interaction. Second, the motor performance of players is captured and processed by applying machine learning algorithms to data collected from inertial sensors worn by players. Artificial intelligence techniques are then used to check the consistency of generated labels, including those regarding the physical activity of players, to automatically recognize annotation errors. Notably, we implemented a full prototype of the proposed system, performing experiments to show its effectiveness in a real-world adoption scenario.
\keywords{Intelligent user interfaces \and Pattern recognition \and Artificial intelligence}
\end{abstract}

\section{Introduction}
\label{sec:introduction}

In the era of Big Data, data annotation activities are very popular, mainly for two reasons: the first is that they can provide a semantic structure to the data, like videos~\cite{Fernandez_2017_ICCV}, images~\cite{8630050}, time-series~\cite{9080613}, and audio file~\cite{7966291}. This allows the fast recovery of specific events from a large amount of information. Secondly, with the ever increasing use of data driven and learning based techniques, like neural networks, deep and shallow learning approaches, the need of labelled data has become fundamental~\cite{9000595}. 
At the state of the art, there are several tools which try to fasten such activities, like LabelImg\footnote{\url{https://pypi.org/project/labelImg/}}, but these have two main drawbacks. The first is that they mainly work on images and the process of scaling the tagging over the frames of a video is quite effort expensive. The second is that the annotation involves a continuous interaction with the user, which needs to manually select the region to annotate, give it a label, save the work, and go on with the next frame.
Obviously, this activity could be feasible for annotating a limited number of images, but it is not the best choice for annotating video events.

Football event annotation has the primary objective of labelling all the events happening in a football match~\cite{sharma2017automatic}. The final goal of the activity is not only the production of a detailed report about the actions of each player (acting in its own role as defender, midfielder, forward, winger, or goalkeeper) during the match, but, in parallel, it aims at collecting team dynamics, in order to analyse forwarding, game construction, and defending tactic movements. Therefore, it is often needed to label a single action under two different points of view: (i) single player actions, expressed as \emph{tag combinations} (e.g., \textsc{midfielder} $\rightarrow$ \textsc{killer pass} $\rightarrow$ \textsc{positive}), and (ii) team actions (i.e., actions which involve more players, such as \textsc{counterattack}). Moreover, in addition to the active plays (i.e., those whose object is the ball), there are plenty of passive plays which need to be labelled as well, in order to properly track the team dynamics, such as double teaming, defending actions, and man-to-man/zone marking. Often, this activity must be done for both teams, in order to provide a full set of tags. 

It is easy to understand that the aforementioned annotation practices involve a considerable workload for those who are in charge of manual annotating the events within a football match. It is indeed estimated that the tagging activities of an entire football match (excluding additional minutes and extra times) may take from 6 to 8 hours for about 2,000 events. 
Annotation of soccer videos is an activity which is very popular in the market, and in the last fifteen years several systems have been proposed in this field. Most approaches are based on a computer vision based analysis over the video stream, with all the pros and cons of this solution. In~\cite{ASSFALG2003285} the authors have proposed a semantic annotation approach for automatically extracting significant clips from a video stream, with the objective of automatically creating a set of highlights of the match. For an analogous purpose, in~\cite{7579433} a rule-based tagging system is proposed, which supports filtering and automatic annotation of soccer events. In~\cite{HOSSEINI2013846} a fuzzy-based reasoning system is proposed for enhancing broadcasting video streams with semantic event annotations. Also, mainly in the last years, many techniques have been invented, which exploit computer graphics, thus giving birth to 3D soccer video understanding, like proposed in~\cite{samanogvr} and in~\cite{MORRA2020100612}. However, the widespread diffusion of learning techniques has enabled a novel generation of approaches and contributions to this research field. Indeed, given the large amount of data yielded by soccer videos, together with the ease of recovering these data, the need for flexible and automatic tools for analysing, validating, testing, and hence predicting/classifying events and potential annotation of soccer videos has arisen. Different works explored deep learning approaches like LSTMs~\cite{sorano2020automatic}, CNNs~\cite{9044306}, GANs~\cite{9316738}, and attention-aware based approaches~\cite{Cioppa2020CVPR,9344896}.

All the cited papers face the annotation problem only using videos (either processed as an online or offline stream). Unfortunately, in many cases, the semantic annotation cannot be precise enough to describe a particular event, mainly due to two reasons:

\begin{itemize}
    \item it is hard to catch information about team plays, like offsides or counterattacks actions. This because such events count on the participation of more than one player, besides needing a deep knowledge about football tactics. Often indeed, the movement of a singular player makes the difference between an action rather than another;
    \item no information about the single movement of the player is used; this is a drawback since, without details about how the player is moving, it is not possible to measure the involvement of a player in a certain action, neither to catch information about those players which are outside the camera frame.
\end{itemize}

In order to address these challenges, in this paper, we propose \emph{FootAPP}, a system that aims at supporting operators in football match annotation. The user is supported by means of three main modules:
\begin{itemize}
    \item a mixed user interface exploiting touch and voice;
    \item a sensor-based football activity recognition module;
    \item an artificial intelligence (AI) algorithm for detecting annotation errors.
\end{itemize}

The work relies on two different types of annotations. The first one is produced by experts who manually annotate the plays happened in a soccer match. The second annotation is automatically extracted from the data stream captured from sensors worn by players. The latter information, besides enriching the manual annotations with further details, is also used to detect annotation errors. In this regard, an AI-based tool is also proposed which compares the event information with those acquired by sensors in order to warn the expert operator in case of conflicting labels.

The main contributions of the paper are:
\begin{itemize}
    \item a twofold soccer videos annotation system is proposed, which helps the synchronization between wide view events and single player activities;
    \item a detector for annotation errors is also developed, which helps to reveal possible conflicts between play events and players activities.
\end{itemize}

The paper is organized as follows. In Section~\ref{sec:overview}, an overview of the proposed system is reported, together with details on the sensor data acquisition process. In Section~\ref{har}, we describe the football activity recognition process applied to wearable sensor data. Section~\ref{errors} details the AI-driven technique for recognizing annotation errors. In Section~\ref{sec:exp}, we describe the event tagging and sensor data infrastructure of FootAPP, together with the database organization, the experimental evaluation, and the achieved results. Section~\ref{conclusion} concludes the paper.

\section{System overview}
\label{sec:overview}
\begin{figure}[h!]
\centering
  \includegraphics[width=\textwidth]{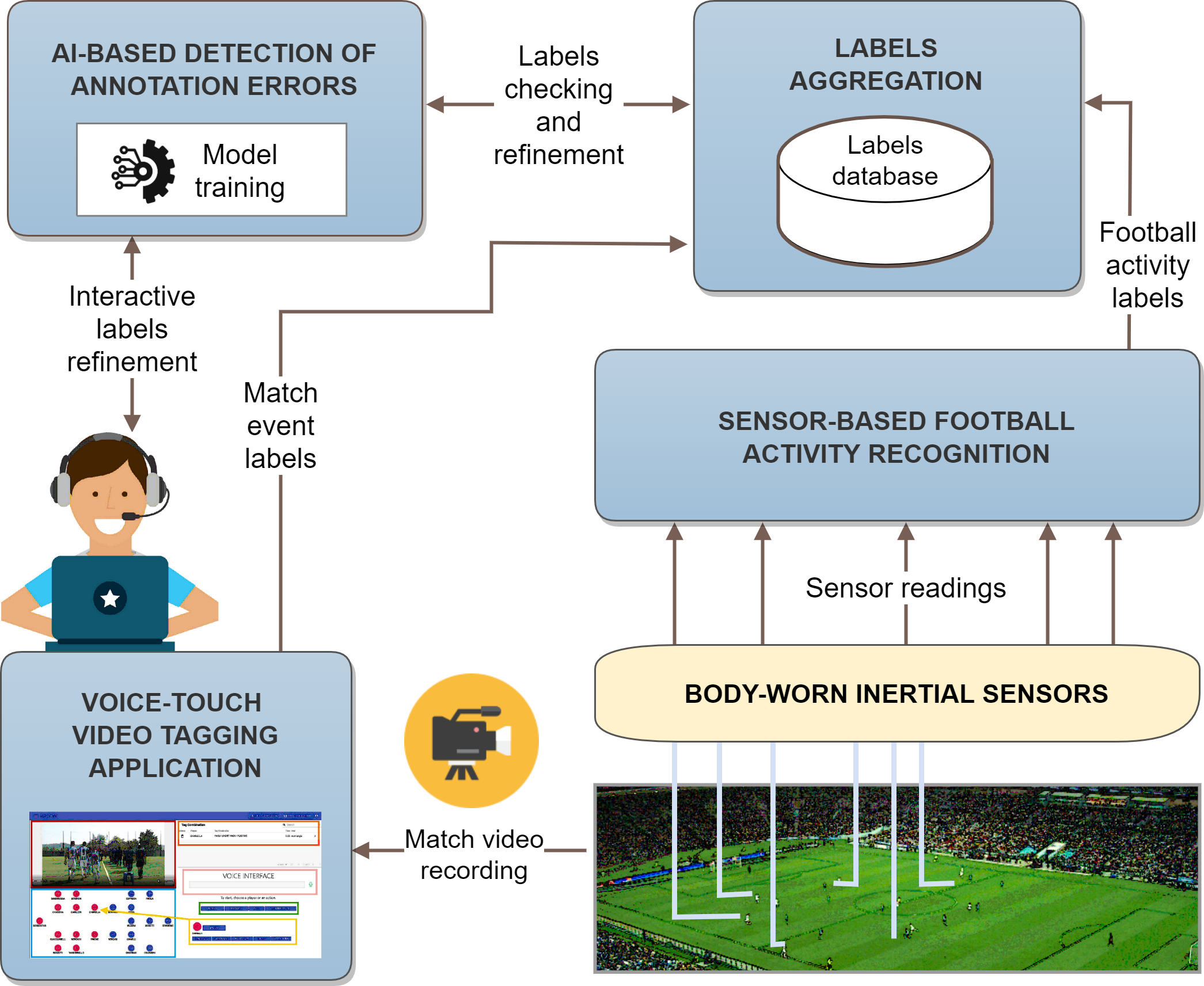}
  \caption{Overall system architecture of FootApp.}
  \label{fig:architecture}
\end{figure}

We hereafter illustrate the overall architecture of the proposed infrastructure. A high-level overview of our architecture is provided in Figure~\ref{fig:architecture}.

As shown in Figure~\ref{fig:architecture}, the main component of FootApp consists of the \textit{Voice-touch Video Tagging Application} \cite{VUI2020}. This application, briefly outlined in Section~\ref{sec:event-tagging-infrastructure}, provides a video player with the recording of the match to analyse and a user-friendly interface for vocal and touch interaction for match tagging. Through it, a human annotator deals with the labeling of football match events in FootApp. The generated labels are then stored in the  \textit{Labels database}.

In addition to the annotation activity, during the game, the involved players are required to wear inertial sensors. Those sensors record all movements made by the players. Then, the \textit{Sensor-based Football Activity Recognition} module of FootApp automatically processes offline the data collected by sensors, thus generating additional labels. The features and capabilities of this module are described in detail in Section~\ref{sec:har}. Roughly, its purpose is to recognize the athletic activity of each player (e.g., running, jumping, falling\ldots) throughout the match, and regardless of whether they are taking an active part in the development of the game. 

Therefore, at the end of the match, the system sends the football activity labels to the \textit{Labels aggregation} module. Specifically, this module owns a shared model of labels, arranged in a hierarchy, and is in charge of integrating the automatically-generated labels with those manually produced by the human annotator. The resulting integrated labels are then stored in the labels database, too. The latter is organized in relational tables and also includes data about football teams, players, matches, leagues, and competitions.

At the highest layer, the system features an \textit{AI-based detection of annotation errors} module, responsible for refining the annotations by discovering -- as its name reveals -- potential annotation errors in stored labels. In particular, it aims at detecting a wide spectrum of errors, including wrong and missing labels. More in detail, it is trained on a large dataset of correct annotations; the trained model is hence used to check novel generated annotations. Whenever the module identifies a possible annotation error, it displays a warning. Such a warning is then manually checked by the human annotator, which exploits the FootApp interface to inspect the video in the time interval affected by the issue and to possibly fix the annotation error. Finally, annotation corrections are applied to the labels database.

\section{Sensor-based football activity recognition}
\label{sec:har}
\label{har}
Physical activity annotations represent an important source of information for assessing the performance of teams and players~\cite{grehaigne1997performance}. Indeed, there is a consensus in the literature on the fact that the performance of sports teams depends not only on tactical and strategic efficiency but also on the motor fitness and athletic skills of players~\cite{carling2008performance}. 

Moreover, in the considered scenario, information about the activity carried out by players can be matched with manually-entered labels to detect inconsistencies due to possible annotation errors. For instance, if the labels database includes a \textit{dummy run}\footnote{A so-called \emph{dummy run} occurs when a player performs an off-the-ball run to create space for his teammate with the ball.} for the player $P$ at time $t$ of the match, but the activity of $P$ at $t$, acquired through the inertial sensors, was \textit{standing still}, the module for the detection of annotation errors presented in Section~\ref{sec:error-detection} may issue an alert. 

\begin{figure}[t!]
\centering
  \includegraphics[width=\textwidth]{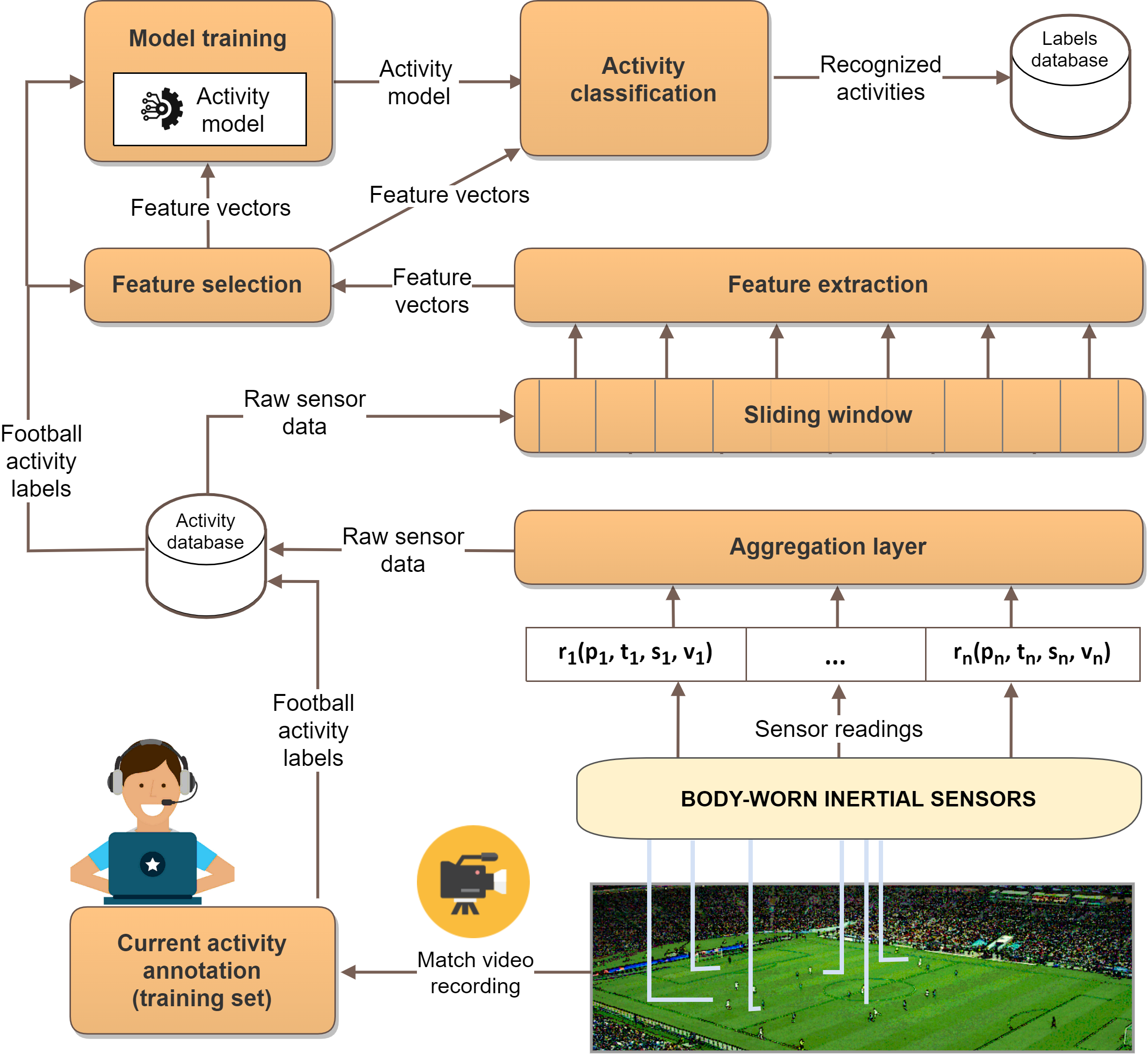}
  \caption{The proposed football activity recognition framework.}
  \label{fig:har}
\end{figure}

Hence, our system includes a specific module designated to automatically annotate the physical activities carried out by players during the match. To recognize the activities, such a module relies on wearable sensors and supervised machine learning. 
Figure~\ref{fig:har} shows the information flow of our football activity recognition framework. As said, it requires players to wear inertial sensing devices on different parts of their bodies during the match. More specifically, each sensing device consists of three sensors: a tri-axial accelerometer, a gyroscope, and a magnetometer. The sensors continuously record data at a given frequency, e.g., 120 Hz. Thus, each sensor provides three streams of readings, computed on the $x$, $y$, and $z$ axis, respectively, for a total of nine measurements at a time. We represent a sensor reading $r$ with the tuple $(p, t, s, v)$, where $p$ is the unique identifier of the player, $t$ is the absolute timestamp of sensor reading, $s$ is the identifier of the sensed value (e.g., \emph{`$x$-axis acceleration of the sensor worn on the left upper arm'}), and $v$ is the sensed value. The data stream acquired from the sensors is then communicated to an \emph{aggregation layer}, which stores the collected data in a relational database and translates the absolute time of sensor readings to the relative time of the match. We also notice that, for our application, the activity recognition stage is performed offline. Consequently, sensor data are not necessarily communicated in real-time to the aggregation layer but may be downloaded in batch mode from the sensing devices. 

Since raw inertial sensor data cannot be directly used for activity recognition~\cite{chen2012sensor,riboni2011cosar}, a \emph{feature extraction} layer extracts feature vectors from the raw data by exploiting a sliding windows approach. Each resulting feature then corresponds to statistics data about some measure (e.g., \emph{`maximum $x$-axis acceleration'}), computed in a single sliding window interval. From the signal produced by a sensor about a given axis during a sliding window of duration $d$, the extraction module computes the following features:

\begin{itemize}
    \item \textit{minimum}, \textit{maximum}, and \textit{average} values, 
    \item \textit{variance}, 
    \item \textit{asymmetry} and \textit{kurtosis},
    \item ten equidistant samples from the \textit{auto-correlation sequence},
    \item the first five \textit{peaks of the Discrete Fourier Transform} (DFT), and the corresponding \textit{frequencies}.
\end{itemize}

Thus, globally, for each signal, the module computes 26 features. Since, as said before, each sensing device produces nine signals, the total number of features for each device is 234. However, when a player wears $n$ sensing devices, the number of features is consequently $234 \cdot n$. Such a large number of features may confuse the classifier and may result in overfitting activity recognition to the training set samples~\cite{liu2005evolving}. To address this problem, we added to our framework an additional \emph{feature selection layer}, to extract a limited set of features that are effective in discriminating among the activity classes. Only the selected features are used to train the classifier and recognize novel instances of activity.

For model training and activity classification, we take a supervised learning approach. As explained in Section~\ref{sec:exp}, in our implementation, we use a Random Forest classifier, but different machine learning algorithms could be used. 
The model training module uses a set of feature vectors that are manually annotated with the current activity of the player to train a model of activities. The machine learning classifier recognizes the current activity class of novel feature vectors using the trained model. Recognized activity instances are finally stored in the labels database. Of course, the set of activities that our system can recognize depends on the activity classes considered in the training set. At the time of writing, we consider 19 activities, including walking, running, lying, jumping. However, the set of activities can be seamlessly extended using a more comprehensive training set.

%\section{Voice-touch video tagging application (Sebastian) \textcolor{red}{da  rifrasare}}
%\label{sec:interface}
%\input{sections/04_voice_touch_interface}

\section{AI-based detection of annotation errors}
\label{sec:error-detection}
\label{errors}
As explained before, match analysis is the task of annotating and analyzing football matches based on the observed events. Each annotation describes an event that occurs during a football match, e.g., ball possession, fouls, or goals. 
Each event can be characterized by the match name, a set of tags, a description label, the time when the event started, the time when it ended, the player being the active part, and his team. 
Moreover, players wear inertial sensors that continuously acquire data about their movements.
Discovering annotation errors and reporting them to the human annotator is useful to improve the quality of the analysis. To this end, the proposed system embeds an automated error detection module based on AI algorithms, as reported in Figure~\ref{fig:architecture}. As explained below, our algorithm relies on Frequent Itemset Mining (FIM)~\cite{luna2019frequent}.  %Several algorithms have been adopted in the field of Match Analysis. Among them, the following algorithm has been selected for being integrated into the AI-based error detection module.

%\paragraph{Frequent Itemset Mining}
In order to identify annotation errors, our algorithm mines a dataset of correct annotation for extracting relationships between match events and activities of the players.
To this end, FIM is one of the most widely used techniques to extract knowledge from data. FIM is currently used for many tasks, and it consists of extracting events, patterns, or items that frequently (or rarely) appear in the data. 
The most used algorithm is \emph{Apriori}~\cite{borgelt2002induction}. It identifies frequent individual items on datasets and extends them to larger itemsets. It is based on the Apriori paradigm, which states that if an itemset is frequent, all its subsets are frequent. Conversely, if an itemset is not frequent, also its supersets are not frequent.
As FIM can extract relationships between elements, it can also be used to infer meaningful relationships between match events and activity instances. 

%In particular, description labels frequently appearing together in a sequence of events and actions may have a specific relationship.
%
A set of temporally close events and activities is represented as a unique entry using a sliding window on the list of annotations acquired during a match. Each entry describes a temporal segment of the match. In this way, the entire match is described as a set of entries. We assume that elements frequently occurring together in entries may be significantly correlated. 
Applying the Apriori algorithm allows the extraction of frequent itemsets representing which combinations of items frequently appear in close temporal proximity. 

Our algorithm applies Apriori to generate a set of \textit{association rules}.
An association rule describes correlations among items belonging to a frequent itemset. Each rule defines an implication, in the form of $X \rightarrow Y$, in which $X$ is a subset called \textit{antecedent itemset} and $Y$ is a subset named \textit{consequent itemset}. If $X$ belongs to an entry, it is probable that also $Y$ appears in the same entry. 
Each rule can be characterized using the following metrics:

\begin{itemize}
  \item \textbf{Support} 
    \[support(X \rightarrow Y) = \frac{number\ of \ entries \ containing \ both\ X\ and \ Y}{total \ number \ of \ entries} .\]
    This measure is used to indicate the percentage of entries in the dataset which respect the rule.
  \vspace{10pt}
  \item \textbf{Confidence}
  \[confidence(X \rightarrow Y) = \frac{number\ of \ entries \ containing \ X \ and \ Y}{total \ number \ of \ entries \ containing \ X} .\]
    This measure estimates the likelihood that $Y$ belongs to an entry given that $X$ belongs to that entry. %Having a confidence of 1 means that $X$ and $Y$ always appear together. %It is the probability of occurrence of consequent given the antecedent.
    \vspace{10pt}
    \item \textbf{Conviction}
    \[conviction(X \rightarrow Y) = \frac{1 - support(Y)}{1 - confidence(X \rightarrow Y)} .\]
     This measure indicates to what extent $Y$ depends on $X$. If $X$ and $Y$ are independent, the value of conviction is 1. % it will be 1.
  %  \vfill
   % \item \textbf{Leverage}
   % \[leverage(X \rightarrow Y) = support(X \rightarrow Y) - support(X) \times support(Y) .\]
   % This measure indicates the difference between the observed frequency X and Y appearing together and the frequency that would be expected if X and Y were independent. When X and Y are independent it is 0. 
\end{itemize}

By applying our algorithm, we can find significant patterns of annotations from the data. For instance, we may derive that, with high confidence, if an entry includes an event of type \emph{pass} by a player $p$, the same entry also includes an activity \emph{kicking} by $p$. That pattern is represented as: \{Pass\} $\rightarrow$ \{Kicking\}. Since some of these patterns can be easily derived from domain knowledge, we complement the association rules automatically extracted from the data with other ones manually defined by domain experts.

Based on the level of support, confidence, and conviction, our system analyses new annotations and reports possible errors to the human annotator. The annotator can choose the level of sensitivity of our system by specifying thresholds for the support, confidence, and conviction values, which would issue alerts of possible errors. 

%The aforementioned mining technique is applied to identify associations between events and activities extracted from both manual annotations and sensor data stored in the database.
%
%Let us point out that some activities cannot be compatible with some events. A typical scenario is when a manual event annotation regards a specific player and, according to its sensors, he is doing an activity that is not compatible with the annotated event. In this situation, the annotation should be an error and needs to be notified to the annotator. 

%\textcolor{red}{Non sarebbe meglio mettere in  questo paragrafo un esempio di incompatibilità, tipo passaggio di un giocatore esplulso etc?}
%

\section{Prototype and experimental evaluation}
\label{sec:exp}
We have developed a full prototype of FootApp. As shown in Figure~\ref{fig:architecture}, our system takes into account two sources of information:

\begin{itemize}
    \item the first one regards the events tagged by the human annotator, which watches the match and, thank to the mixed touch-vocal User Interface, tags all the events happening in the scene;
    \item the second source provides players' activity annotations obtained from the application of AI techniques to the data acquired by body-worn sensors. %This information are strictly related to the movement parameters of the players on the pitch. 
\end{itemize}

\begin{figure}[t!]
\centering
  \includegraphics[width=\textwidth]{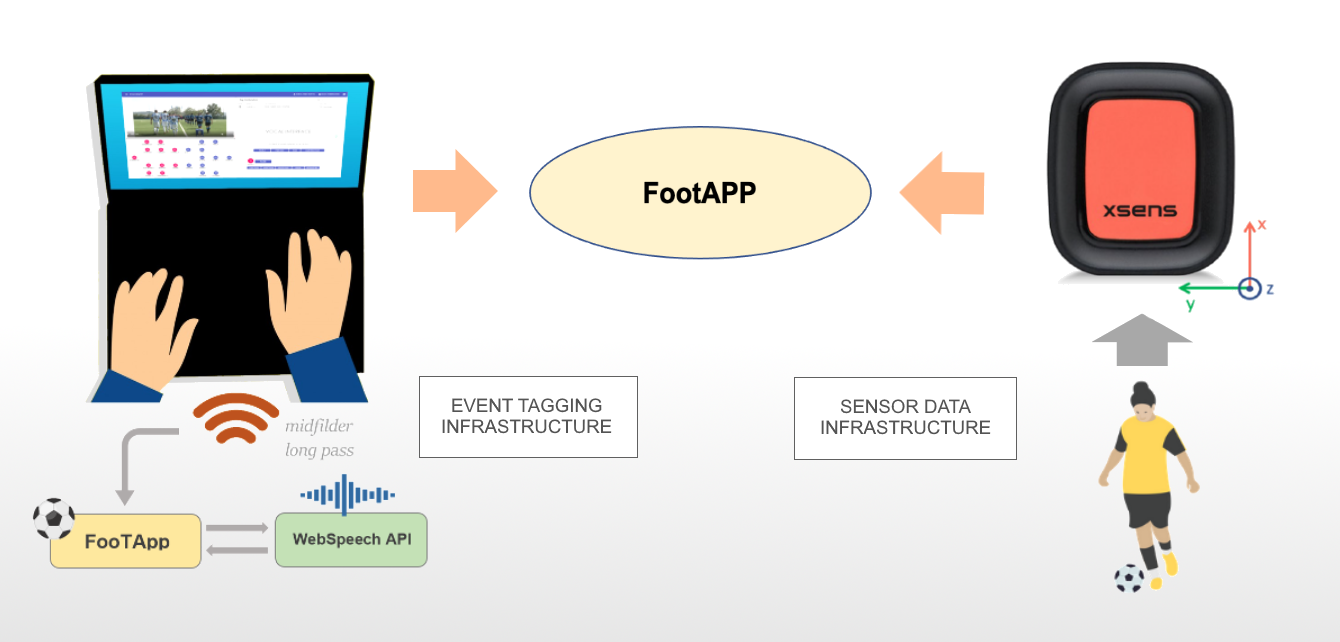}
  \caption{Event tagging and Sensor Data infrastructures.}
  \label{fig:inras}
\end{figure}

These infrastructures act as shown in Figure~\ref{fig:inras}. The labelled contextual data are organized into a Relational database, which gathers and categorizes all the information.
The match events are acquired based on the process described in \cite{VUI2020}, in which a vocal interface is exploited for fastening the annotation of all the events in the scene.

\subsection{Event tagging infrastructure}
\label{sec:event-tagging-infrastructure}
The event tagging infrastructure interacts with the user by means of a twofold modality. Details are reported in our previous work \cite{VUI2020}.
The system exploits a \emph{Voice User Interface} (VUI), integrated with a touch interface, to enhance the user experience and usability of the FootApp web application. 

The underlying idea that guided its design consists in reducing the time needed for tagging matches, with a twofold advantage: i) first, the exploitation of a multi-modal interface, which combines the benefits of both voice interaction (e.g., for longer and more articulated commands) and touch interaction (for the direct player selection or click on a suggested tag, etc.); ii) a greater accessibility of the application for people who have difficulty in using traditional interfaces (mouse and keyboard), so as to choose the most suited mode to them, while maintaining a high level of efficiency in the execution of the task.

\begin{figure}[t!]
\centering
  \includegraphics[width=\textwidth]{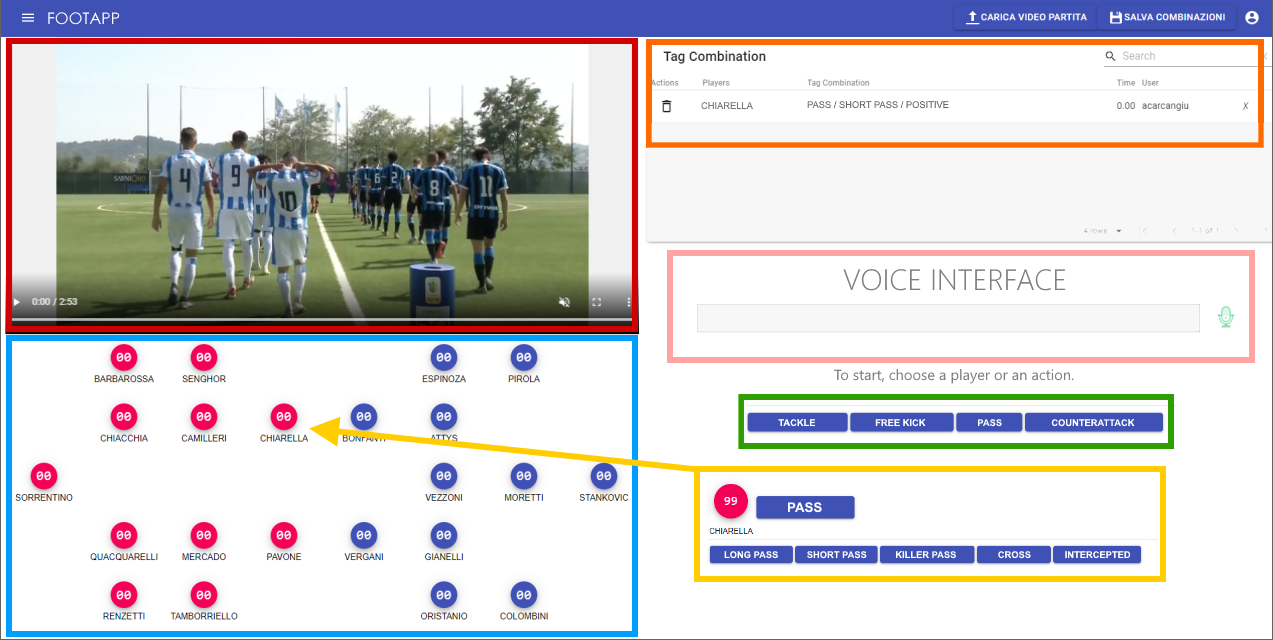}
  \caption{The Vocal User Interface of FootAPP.}
  \label{fig:teaser}
\end{figure}

\paragraph{User Interface.}
Currently, FootApp GUI is organized as shown in Fig.~\ref{fig:teaser}. We describe its components in what follows:

\begin{itemize}
    \item the top left block (contoured in red in Fig.~\ref{fig:teaser}) contains a video player, in which the match is reproduced; directional arrows ease the forward and backward skips of the video;
    \item the bottom left block (contoured in cyan) shows a virtual field in which the team lineups and shape are shown; the buttons with player numbers are clickable, to be inserted in a tag combination;
    \item the top right block (contoured in orange) contains a summary of the tag combination records already registered for the current match;
    \item the middle right block (contoured in pink) contains the voice interface, which automatically activates when the user casts a command;
    \item the block contoured in green shows the keyboard which activates as the user selects a player from the virtual field or spells his number (first level tags); 
    \item the block contoured in yellow contains additional details associated to the selected event. This block pops up when a main event is activated, and could potentially provide a large number of second/third level options.
\end{itemize}

\paragraph{VUI integration and empirical tests.}
We developed the voice user interface by exploiting the Web Speech API \cite{adorf2013web}, which defines a complex interface (\textit{SpeechRecognition}) that provides a set of methods for transforming speech into text. 
The implemented VUI structure is shown in Fig.~\ref{fig:teaser}. From preliminary empirical tests conducted with professional annotators of football match events, we noticed that total migration to a VUI would have partially obtained the expected benefits. Then, we adapted the original interface to enable a combined tagging mode (voice and touch), resulting in a significant reduction for annotating a single football match, empirically estimated at around -25\% than the time required by a traditional system. In particular, the introduction of touch functionalities, coupled with the vocal interface, had a positive effect for two main reasons. First, the processing time of long voice commands (which, despite the robustness of the exploited API, can last for a few seconds) could negatively impact the annotation performance. Second, it is quicker to execute some tagging patterns (e.g., the selection of the player from the virtual field in the bottom left block of Figure~\ref{fig:teaser}) in touch modality than in vocal mode. On the contrary, other tagging practices (as an example, finding second/third level tags among many options) can obtain real benefits from the use of the voice interface.

\subsection{Sensor data infrastructure}
In our prototype, for the acquisition of inertial data from body-worn sensors, we used the \textsc{XSens Dot Sensor} instruments, shown in Figure~\ref{fig:inras}. 
%
%\textcolor{red}{rivedere ed eventualmente correggere}
The sensors detect and store the following data at 120 Hz frequency:

\begin{itemize}
    \item Orientation, computed by means the Euler's angles on the $x$, $y$ and $z$ axes (in degrees);
    \item Acceleration on $x$, $y$ and $z$ axes, both with and without gravity (in $m/s^2$);
    \item Angular speed on $x$, $y$ and $z$ axes (in degrees$/s^2$);
    \item Magnetic field on  $x$, $y$ and $z$ axes, comprising of normalization factor.
\end{itemize}

Such data are then downloaded (in offline mode) into a \texttt{csv} file by means of an application running on an Android device. In this phase, a synchronization phase is needed, given the time difference between the Android application, which is synchronized to the international time, and the sensors which counts the time since the moment it is powered on. This operation is quite simple and it is done by registering the moment in which a specific sensor is powered on, thus synchronizing this timestamp with the international measure followed by the Android application. 
The prototype, which is implemented in Python, follows three main steps:
\begin{itemize}
    \item Data acquisition and feature extraction;
    \item Classifier training and activity prediction;
    \item Prediction storing into the database.
\end{itemize}

In Section~\ref{har}, the 234 features extraction process is described. Given the high number of features, in order to avoid overfitting, a feature selection process takes place, in which the 30 best features are extracted. This operation is done using a \textit{chi-squared} based selection process which removes all the features but the best $k$, with $k=30$ \cite{haryanto2018influence}. The selected features feed a %SVM 
machine learning classifier. In the current implementation, we use a Random Forest classifier. %, with a linear kernel. 
The prediction obtained for each activity is then inserted into the database of annotations (developed in PostgreSQL), which is described below.

\subsection{Database description}
\begin{figure}[t!]
\centering
  \includegraphics[width=\textwidth]{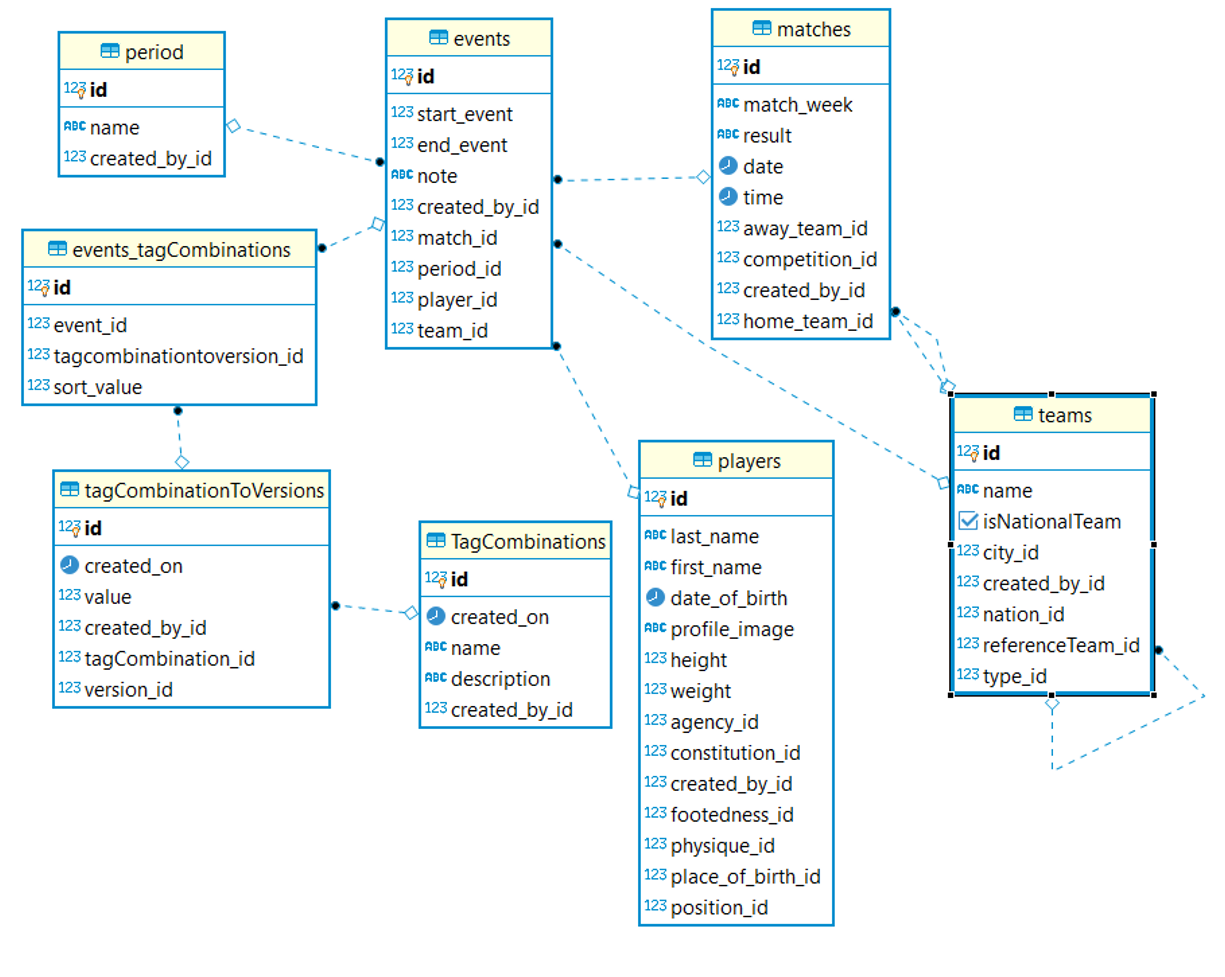}
  \caption{The \textit{XSens Dot Platform} used for registering players' body movement on the pitch.}
  \label{fig:db}
\end{figure}

The database consists of about 90 tables, which allow the management of the entire set of data, including information about players, teams, coaches, matches and predicted activities too. Part of the database diagram is shown in Figure~\ref{fig:db}. 
In the table \textsc{events} all the activities of the single players are recorded. The tables contain several foreign keys which allow the recovery of the information regarding the player (\textsc{player\_id}), the match (\textsc{match\_id}), the half of the match (\textsc{period\_id}) and the team of the player (\textsc{team\_id}). The single activity is described into the table \textsc{TagCombinations} which contains its name and description.

For the most common queries, prepared statement have been defined like the one in the following, in which by defining \textsc{match\_id}, \textsc{period\_id} and \textsc{player\_id}, all the events happened in a specific match, in a specific period and involving a specific player are returned in chronological order.
\begin{verbatim}
  SELECT start_event, end_event, "TagCombinations".name
  FROM "TagCombinations"
  JOIN "tagCombinationToVersions" 
  ON "tagCombination_id" = "TagCombinations".id
  JOIN "event_tagCombinations" 
  ON "tagcombinationtoversion_id" = "tagCombinationToVersion".id
  JOIN events on events.id = "events_tagCombinations".event_id
  JOIN players on players.id = events.players_id
  WHERE match_id = ? AND player_id = ? AND period_id = ?
  ORDER BY start_event
\end{verbatim}

\subsection{Experimental results}
In the following, we report experimental results on our modules for activity recognition and for detection of annotation errors, respectively.

\subsubsection{Football activity recognition}
In order to evaluate the effectiveness of our activity recognition method presented in Section~\ref{sec:har}, we have used a large dataset of physical and sport activities. The dataset considers 19 activities, including walking (at different speeds), running, lying, jumping, standing. For each activity, the data have been acquired from 8 different subjects (4 males and 4 females, aged 20 to 30) for 5 minutes per subject. Hence, the total temporal duration of the dataset is 760 minutes. Sensor data have been acquired at 25 Hz sampling frequency. Five sensing devices were worn by each subject: one on each knee, one on each wrist, and one on the chest. Each sensing device included one tri-axial accelerometer, one tri-axial gyroscope, and one tri-axial magnetometer. The data have been acquired in naturalistic conditions. The dataset is described in detail in~\cite{altun2010comparative}.

We have developed the algorithms presented in Section~\ref{sec:har} in python, using the \emph{sklearn} machine learning libraries for implementing the feature selection and classification functions. We have used a \emph{leave-one-subject-out} cross-validation approach, using one subject's data for the test set and the other subjects' data for training. For the classifier, we have used two algorithms: Random Forest~\cite{verikas2011mining}, and Support Vector Machines~\cite{shawe2000support}, since in the literature they are considered among the most effective classifiers for sensor-based activity recognition~\cite{chen2012sensor}.

\begin{table}[t!]
  \centering
  \caption{Precision, recall, and F-score for each activity class using Random Forest (left) and Support Vector Machines (right).}
    \begin{tabular}{l|l|l|lrl|l|l|l}
    Precision & Recall & F-score & Activity &       & Precision & Recall & F-score & Activity \\
\cmidrule{1-4}\cmidrule{6-9}    \rowcolor[rgb]{ .949,  .949,  .949} 1     & 1     & 1     & A1    & \cellcolor[rgb]{ 1,  1,  1} & 1     & 1     & 1     & A1 \\
    0.811 & 0.627 & 0.707 & A2    &       & 0.403 & 0.755 & 0.53  & A2 \\
    \rowcolor[rgb]{ .949,  .949,  .949} 1     & 0.875 & 0.933 & A3    & \cellcolor[rgb]{ 1,  1,  1} & 1     & 0.992 & 0.996 & A3 \\
    0.889 & 1     & 0.941 & A4    &       & 0.996 & 1     & 0.998 & A4 \\
    \rowcolor[rgb]{ .949,  .949,  .949} 0.699 & 0.915 & 0.792 & A5    & \cellcolor[rgb]{ 1,  1,  1} & 0.666 & 0.888 & 0.761 & A5 \\
    0.95  & 0.902 & 0.925 & A6    &       & 0.805 & 0.863 & 833   & A6 \\
    \rowcolor[rgb]{ .949,  .949,  .949} 0.623 & 0.729 & 0.672 & A7    & \cellcolor[rgb]{ 1,  1,  1} & 0.504 & 0.263 & 0.346 & A7 \\
    0.747 & 0.795 & 0.77  & A8    &       & 0.706 & 0.34  & 0.459 & A8 \\
    \rowcolor[rgb]{ .949,  .949,  .949} 0.57  & 0.625 & 0.596 & A9    & \cellcolor[rgb]{ 1,  1,  1} & 0.584 & 0.615 & 0.599 & A9 \\
    0.57  & 0.631 & 0.599 & A10   &       & 0.575 & 0.744 & 0.649 & A10 \\
    \rowcolor[rgb]{ .949,  .949,  .949} 0.791 & 0.425 & 0.553 & A11   & \cellcolor[rgb]{ 1,  1,  1} & 0.792 & 0.61  & 0.689 & A11 \\
    0.996 & 1     & 0.998 & A12   &       & 1     & 1     & 1     & A12 \\
    \rowcolor[rgb]{ .949,  .949,  .949} 0.867 & 0.831 & 0.849 & A13   & \cellcolor[rgb]{ 1,  1,  1} & 0.86  & 0.654 & 0.743 & A13 \\
    0.768 & 0.746 & 0.757 & A14   &       & 0.719 & 0.625 & 0.669 & A14 \\
    \rowcolor[rgb]{ .949,  .949,  .949} 0.966 & 1     & 0.983 & A15   & \cellcolor[rgb]{ 1,  1,  1} & 1     & 1     & 1     & A15 \\
    0.998 & 0.965 & 0.981 & A16   &       & 1     & 1     & 1     & A16 \\
    \rowcolor[rgb]{ .949,  .949,  .949} 1     & 1     & 1     & A17   & \cellcolor[rgb]{ 1,  1,  1} & 0.996 & 1     & 0.998 & A17 \\
    1     & 0.965 & 0.982 & A18   &       & 0.996 & 992   & 0.994 & A18 \\
    \rowcolor[rgb]{ .949,  .949,  .949} 0.851 & 0.931 & 0.89  & A19   & \cellcolor[rgb]{ 1,  1,  1} & 0.873 & 0.898 & 0.934 & A19 \\
\cmidrule{1-4}\cmidrule{6-9}    \textbf{0.847} & \textbf{0.84}  & \textbf{0.838} & Average &       & \textbf{0.82}  & \textbf{0.803} & \textbf{0.8}   & Average \\
    \end{tabular}%
  \label{tab:PRREF}%
\end{table}%

Results are shown in Table~\ref{tab:PRREF}. We use the standard metrics of precision, recall, and F-score, the latter being the harmonic mean of precision and recall. Overall, our results are in line with the the best results achieved by state-of-the-art methods on the same dataset~\cite{altun2010comparative}. By closely inspecting the results shown in the confusion matrix, reported in Table~\ref{tab:randomForest} for the Random forest classifier and in Table~\ref{tab:SVM} for the Support Vector Machines classifier, respectively, we can notice that a few activity classes (in particular, A9, A10, and A11) achieve rather low values of F-score. Those activity classes represent different kinds of walking, which are difficult to distinguish based on sensor data. By merging those classes in a unique class, we could improve the overall recognition rate without disrupting the data utility for our application. 

\begin{table}[t!]
  \centering
  \caption{Confusion matrix obtained with the Random forest classifier}
  \scalebox{0.63}{
       \begin{tabular}{c|rrrrrrrrrrrrrrrrrrr}
    Activity & \multicolumn{19}{l}{Classified as} \\
\cmidrule{2-20}          & \multicolumn{1}{c}{A1} & \multicolumn{1}{c}{A2} & \multicolumn{1}{c}{A3} & \multicolumn{1}{c}{A4} & \multicolumn{1}{c}{A5} & \multicolumn{1}{c}{A6} & \multicolumn{1}{c}{A7} & \multicolumn{1}{c}{A8} & \multicolumn{1}{c}{A9} & \multicolumn{1}{c}{A10} & \multicolumn{1}{c}{A11} & \multicolumn{1}{c}{A12} & \multicolumn{1}{c}{A13} & \multicolumn{1}{c}{A14} & \multicolumn{1}{c}{A15} & \multicolumn{1}{c}{A16} & \multicolumn{1}{c}{A17} & \multicolumn{1}{c}{A18} & \multicolumn{1}{c}{A19} \\
    \midrule
    A1    & 480   & 0     & 0     & 0     & 0     & 0     & 0     & 0     & 0     & 0     & 0     & 0     & 0     & 0     & 0     & 0     & 0     & 0     & 0 \\
    A2    & 0     & 301   & 0     & 0     & 0     & 0     & 155   & 24    & 0     & 0     & 0     & 0     & 0     & 0     & 0     & 0     & 0     & 0     & 0 \\
    A3    & 0     & 0     & 420   & 60    & 0     & 0     & 0     & 0     & 0     & 0     & 0     & 0     & 0     & 0     & 0     & 0     & 0     & 0     & 0 \\
    A4    & 0     & 0     & 0     & 480   & 0     & 0     & 0     & 0     & 0     & 0     & 0     & 0     & 0     & 0     & 0     & 0     & 0     & 0     & 0 \\
    A5    & 0     & 0     & 0     & 0     & 439   & 3     & 0     & 2     & 0     & 0     & 5     & 0     & 21    & 10    & 0     & 0     & 0     & 0     & 0 \\
    A6    & 0     & 0     & 0     & 0     & 11    & 433   & 0     & 8     & 4     & 0     & 0     & 0     & 1     & 0     & 0     & 1     & 0     & 0     & 22 \\
    A7    & 0     & 54    & 0     & 0     & 0     & 0     & 349   & 76    & 0     & 0     & 0     & 0     & 0     & 0     & 0     & 0     & 0     & 0     & 0 \\
    A8    & 0     & 16    & 0     & 0     & 2     & 1     & 56    & 381   & 14    & 0     & 0     & 0     & 0     & 0     & 0     & 0     & 0     & 0     & 9 \\
    A9    & 0     & 0     & 0     & 0     & 29    & 9     & 0     & 1     & 300   & 120   & 11    & 0     & 3     & 0     & 0     & 0     & 0     & 0     & 7 \\
    A10   & 0     & 0     & 0     & 0     & 0     & 0     & 0     & 0     & 158   & 303   & 18    & 0     & 0     & 1     & 0     & 0     & 0     & 0     & 0 \\
    A11   & 0     & 0     & 0     & 0     & 52    & 0     & 0     & 0     & 16    & 105   & 204   & 0     & 12    & 91    & 0     & 0     & 0     & 0     & 0 \\
    A12   & 0     & 0     & 0     & 0     & 0     & 0     & 0     & 0     & 0     & 0     & 0     & 480   & 0     & 0     & 0     & 0     & 0     & 0     & 0 \\
    A13   & 0     & 0     & 0     & 0     & 51    & 8     & 0     & 0     & 7     & 4     & 1     & 0     & 399   & 6     & 0     & 0     & 0     & 0     & 4 \\
    A14   & 0     & 0     & 0     & 0     & 39    & 0     & 0     & 1     & 20    & 0     & 19    & 0     & 24    & 358   & 0     & 0     & 0     & 0     & 19 \\
    A15   & 0     & 0     & 0     & 0     & 0     & 0     & 0     & 0     & 0     & 0     & 0     & 0     & 0     & 0     & 480   & 0     & 0     & 0     & 0 \\
    A16   & 0     & 0     & 0     & 0     & 0     & 0     & 0     & 0     & 0     & 0     & 0     & 0     & 0     & 0     & 17    & 463   & 0     & 0     & 0 \\
    A17   & 0     & 0     & 0     & 0     & 0     & 0     & 0     & 0     & 0     & 0     & 0     & 0     & 0     & 0     & 0     & 0     & 480   & 0     & 0 \\
    A18   & 0     & 0     & 0     & 0     & 0     & 0     & 0     & 0     & 0     & 0     & 0     & 0     & 0     & 0     & 0     & 0     & 0     & 463   & 17 \\
    A19   & 0     & 0     & 0     & 0     & 5     & 2     & 0     & 17    & 7     & 0     & 0     & 2     & 0     & 0     & 0     & 0     & 0     & 0     & 447 \\
    \end{tabular}%
    }
  \label{tab:randomForest}%
\end{table}%

\begin{table}[t!]
  \centering
  \caption{Confusion matrix obtained with the SVM classifier}
  \scalebox{0.63}{
    \begin{tabular}{c|rrrrrrrrrrrrrrrrrrr}
    Activity & \multicolumn{19}{l}{Classified as} \\
\cmidrule{2-20}          & \multicolumn{1}{c}{A1} & \multicolumn{1}{c}{A2} & \multicolumn{1}{c}{A3} & \multicolumn{1}{c}{A4} & \multicolumn{1}{c}{A5} & \multicolumn{1}{c}{A6} & \multicolumn{1}{c}{A7} & \multicolumn{1}{c}{A8} & \multicolumn{1}{c}{A9} & \multicolumn{1}{c}{A10} & \multicolumn{1}{c}{A11} & \multicolumn{1}{c}{A12} & \multicolumn{1}{c}{A13} & \multicolumn{1}{c}{A14} & \multicolumn{1}{c}{A15} & \multicolumn{1}{c}{A16} & \multicolumn{1}{c}{A17} & \multicolumn{1}{c}{A18} & \multicolumn{1}{c}{A19} \\
    \midrule
    A1    & 480   & 0     & 0     & 0     & 0     & 0     & 0     & 0     & 0     & 0     & 0     & 0     & 0     & 0     & 0     & 0     & 0     & 0     & 0 \\
    A2    & 0     & 372   & 0     & 0     & 0     & 0     & 79    & 29    & 0     & 0     & 0     & 0     & 0     & 0     & 0     & 0     & 0     & 0     & 0 \\
    A3    & 0     & 0     & 476   & 2     & 0     & 0     & 0     & 0     & 0     & 0     & 0     & 0     & 0     & 0     & 0     & 0     & 2     & 0     & 0 \\
    A4    & 0     & 0     & 0     & 480   & 0     & 0     & 0     & 0     & 0     & 0     & 0     & 0     & 0     & 0     & 0     & 0     & 0     & 0     & 0 \\
    A5    & 0     & 0     & 0     & 0     & 426   & 12    & 0     & 0     & 0     & 0     & 0     & 0     & 22    & 20    & 0     & 0     & 0     & 0     & 0 \\
    A6    & 0     & 0     & 0     & 0     & 30    & 414   & 0     & 17    & 17    & 0     & 0     & 0     & 1     & 0     & 0     & 0     & 0     & 0     & 1 \\
    A7    & 0     & 352   & 0     & 0     & 0     & 0     & 126   & 1     & 0     & 0     & 0     & 0     & 0     & 0     & 0     & 0     & 0     & 0     & 0 \\
    A8    & 0     & 200   & 0     & 0     & 0     & 43    & 45    & 163   & 20    & 0     & 0     & 0     & 1     & 0     & 0     & 0     & 0     & 0     & 7 \\
    A9    & 0     & 0     & 0     & 0     & 2     & 13    & 0     & 1     & 295   & 144   & 10    & 0     & 1     & 14    & 0     & 0     & 0     & 0     & 0 \\
    A10   & 0     & 0     & 0     & 0     & 0     & 0     & 0     & 0     & 121   & 357   & 2     & 0     & 0     & 0     & 0     & 0     & 0     & 0     & 0 \\
    A11   & 0     & 0     & 0     & 0     & 10    & 0     & 0     & 0     & 14    & 96    & 293   & 0     & 1     & 66    & 0     & 0     & 0     & 0     & 0 \\
    A12   & 0     & 0     & 0     & 0     & 0     & 0     & 0     & 0     & 0     & 0     & 0     & 480   & 0     & 0     & 0     & 0     & 0     & 0     & 0 \\
    A13   & 0     & 0     & 0     & 0     & 83    & 14    & 0     & 4     & 24    & 20    & 4     & 0     & 314   & 17    & 0     & 0     & 0     & 0     & 0 \\
    A14   & 0     & 0     & 0     & 0     & 87    & 0     & 0     & 0     & 7     & 2     & 61    & 0     & 23    & 300   & 0     & 0     & 0     & 0     & 0 \\
    A15   & 0     & 0     & 0     & 0     & 0     & 0     & 0     & 0     & 0     & 0     & 0     & 0     & 0     & 0     & 480   & 0     & 0     & 0     & 0 \\
    A16   & 0     & 0     & 0     & 0     & 0     & 0     & 0     & 0     & 0     & 0     & 0     & 0     & 0     & 0     & 0     & 480   & 0     & 0     & 0 \\
    A17   & 0     & 0     & 0     & 0     & 0     & 0     & 0     & 0     & 0     & 0     & 0     & 0     & 0     & 0     & 0     & 0     & 480   & 0     & 0 \\
    A18   & 0     & 0     & 0     & 0     & 0     & 0     & 0     & 0     & 0     & 0     & 0     & 0     & 0     & 0     & 0     & 0     & 0     & 476   & 4 \\
    A19   & 0     & 0     & 0     & 0     & 2     & 18    & 0     & 16    & 2     & 2     & 0     & 0     & 2     & 0     & 0     & 0     & 0     & 2     & 431 \\
    \end{tabular}%
}
  \label{tab:SVM}%
\end{table}%

\subsubsection{Detection of annotation errors}
We have used the Python language to implement and experiment the frequent itemset mining technique explained in Section~\ref{sec:error-detection}. The libraries that we used are $pandas$, $numpy$ and $mlxtend$.
The dataset of the manually annotated events is structured in the following way:
\begin{itemize}
    \item \textit{Episode}: an integer that identifies the event or activity in the match;
    \item \textit{Match}: a string that identifies the name of the match;
    \item \textit{Team}: a string that identifies the name of the team having an active part in the episode;
    \item \textit{Start}: the time (in the format minutes:seconds) of the episode start;
    \item \textit{End}: the time (in the format minutes:seconds) of the episode end;
    \item \textit{Half}: an integer that indicates in which half of the match the episode occurred;
    \item \textit{Description}: a string that describes the type of the episode;
    \item \textit{Tags}: a list of strings that characterize the episode and its details;
    \item \textit{Player}: a string that indicates the name of the player who had an active part in the episode;
    \item \textit{Notes}: a string that contains comments or notes about the episode.
\end{itemize}

For the sake of annotation error detection, we decided to reduce the modelling of the episode to its Description field, since it is the most representative data.
Using the Apriori algorithm, we mined the frequent itemsets of episodes from a dataset of matches labeled by professional annotators. The dataset included annotations from 64 football matches: among them, 58 were matches of the Campionato Primavera 1 competition (i.e., the premier youth soccer league in Italy), while 6 were international under-17 matches. Totally, the dataset included 100,892 annotations. Then, the itemsets were used to construct the association rules using the method provided by the $mlxtend$ library. 
Each rule was analysed by a domain expert to check whether it was representative of a real-world behaviour of football match events and activities. Other rules were manually added to specify behaviours useful in the domain but not captured by our data-driven technique. %They are useful because they are able to better describe the behaviour of the game. 
As explained in Section~\ref{sec:error-detection}, manually added rules have a qualitative estimation of their confidence level, but lack values in the other metrics (i.e., support and conviction). 
Table 4 shows a sample of the mined rules.

\begin{table}[t!]
  \caption{Association rules for annotation error detection.}

\begin{tabular}{l|c|c|c}
\hline
\multicolumn{1}{c}{\textbf{Rule}}                       & \textbf{Confidence} & \textbf{Support} & \textbf{Conviction} \\ \hline

\{Reception, Construction\} $\rightarrow$ \{Pass\}       & 0.90036             & 0.01383            & 1.95564               \\ \hline
\{Unmarking\} $\rightarrow$ \{Pass\}                      & 0.8845              & 0.04883            & 1.6879                \\ \hline
\{Interception, Reception\} $\rightarrow$ \{Pass\}        & 0.8783              & 0.0129             & 1.60229               \\ \hline
\{Reception, Unmarking\} $\rightarrow$ \{Pass\}           & 0.8667              & 0.0108             & 1.4623                \\ \hline
\{Interception\} $\rightarrow$ \{Pass\}                   & 0.86615             & 0.1682             & 1.4557                \\ \hline
\{Reception\} $\rightarrow$ \{Pass\}                      & 0.8654              & 0.0997             & 1.4476                \\ \hline
\{Offensive moves\} $\rightarrow$ \{Pass\}                & 0.8591              & 0.03414            & 1.3837                \\ \hline
\{Lose the ball\} $\rightarrow$ \{Recover the ball\}      & 0.62644             & 0.03961            & 2.503                 \\ \hline
\{Transmission\} $\rightarrow$ \{Construction\}           & 0.52612             & 0.02495            & 1.7553                \\ \hline
\{Shielding\} $\rightarrow$ \{Reception\}                 & 0.47099             & 0.0087             & 1.672                 \\ \hline
\{Cross Interception\} $\rightarrow$ \{Pass refinement\} & 0.4658              & 0.0201             & 1.6407                \\ \hline
\{Offensive moves\} $\rightarrow$ \{Reception\}           & 0.4335              & 0.01722            & 1.561                 \\ \hline
\{Pass, Offensive moves\} $\rightarrow$ \{Reception\}     & 0.432261            & 0.01475            & 1.5584                \\ \hline
\{Shielding\} $\rightarrow$ \{Defensive duel\}            & 0.4251              & 0.0078             & 1.1352                \\ \hline
\{Cross Interception\} $\rightarrow$ \{Line Attack\}      & 0.3020              & 0.013              & 1.227                 \\ \hline
\{Pass\} $\rightarrow$ \{Kicking\}                        & High                & \multicolumn{2}{c}{Manually defined rule} \\ \hline
\{Header\} $\rightarrow$ \{Jumping\}                      & Medium              & \multicolumn{2}{c}{Manually defined rule} \\ \hline
\{Offensive moves\} $\rightarrow$ \{Running\}             & Medium              & \multicolumn{2}{c}{Manually defined rule} \\ \hline
\{Construction\} $\rightarrow$ \{Running\}                & Low                 & \multicolumn{2}{c}{Manually defined rule} \\ \hline
\{Shielding\} $\rightarrow$ \{Standing still\}            & Low                 & \multicolumn{2}{c}{Manually defined rule} \\ \hline
\end{tabular}
\end{table}

Those association rules represent the normal trend of a game. If an event does not match these rules, we assume it could be an error. 
For instance, the first rule has obtained a Confidence value of 0.90. This measure means that $Reception$ and $Construction$ appear in the 90\% of events that have $Pass$ as a neighbour event in the temporal window. That rule has a support value of 0.013; hence, the itemset containing $\{Reception,$ $Construction,$ $Pass\}$ is not very frequent \emph{per se}, but this is not surprising considering the large number of episode types and the high variability of plays in football. The conviction value of that rule is 1.955, meaning that the two itemsets are significantly correlated.

%For example, if an event registered at a specific time says that the player $x$ is $dribbling$ but his sensors say that he is $standing still$ then, that event may be an error committed by the annotator. This is because the event $dribbling$ is not associated with the activity $standing still$, so there is an anomaly.
%However, if an event says that player $y$ is $Passing$ the ball and his sensors say that $y$ is $kicking$, then this event makes complete sense and it is not an error. This is because the action $kicking$ and the event $pass$ are compatible and are associated in a rule. 

\section{Conclusion}
\label{conclusion}
In this paper, we have presented FootApp, an innovative interface for football match event annotation. FootApp relies on a mixed interface exploiting vocal and touch interaction. Thanks to the use of body-worn inertial sensors and supervised learning, our system can automatically acquire labels regarding the activity carried out by players during the match. An AI module is in charge of detecting inconsistencies in the labels, including those describing the players' activity, for recognizing possible annotation errors. We have implemented a full prototype of our system and experimented with the system. The results indicate the effectiveness of our algorithms.

\begin{acknowledgements}
This work was partially funded by the POR FESR Sardegna 2014-2020 project ``MISTER: Match Information System and Technologies for the Evaluation of the Performance''.
\end{acknowledgements}

%\section*{Conflict of interest}

%The authors declare that they have no conflict of interest.


\bibliographystyle{spmpsci}
\bibliography{refs.bib}
%\begin{thebibliography}{}
% etc
%\end{thebibliography}

\end{document}